# Intensity-Dependent Enhancement of Saturable Absorption in PbS-Au$_4$ Nanohybrid Composites: Evidence for Resonant Energy Transfer by Auger Recombination

Hendry I. Elim and Wei Ji[*]

*Department of Physics, National University of Singapore, 2 Science Drive 3, Singapore 117542*

Jian Yang and Jim Yang Lee

*Department of Chemical and Biomolecular Engineering, National University of Singapore, 4 Engineering Drive 3, Singapore 117576*

*Abstract*

Intensity-dependent enhancement of saturable absorption in a film of PbS-Au$_4$ nanohybrid composites has been observed by femtosecond time-resolved transient absorption measurement at 780 nm. The nonlinear absorption coefficient ($\alpha_2$) of saturable absorption in PbS-Au$_4$ nanohybrid composites is found to be dependent on excitation irradiance and it is determined to be -2.9 cm/GW at 78 GW/cm$^2$, an enhancement of nearly fourfold in comparison with that of pure PbS quantum dots (QDs). The enhancement is attributed to excitation of surface plasmon by resonant energy transfer between PbS QDs and Au nanoparticles through Auger recombination.

---

[*] E-mail: phyjiwei@nus.edu.sg

Metal-semiconductor and semiconductor-semiconductor nanohybrid composites or core/shell nanostructures have attracted increasing attention because of their optical properties, photocatalytic activities, and ultrafast carrier dynamics.[1-7] Understanding of such systems is of direct relevance to future development of electronic and photonic devices. Energy transfer of charge carriers excited by two-photon absorption in CdS has been observed to the surface plasmon of Au in Au-CdS core/shell nanocomposites.[5] Another nanohybrid composite, CdSe-Au nanodumbbell, has been reported to exhibit quenching of photoluminescence (PL) as gold tip is grown onto CdSe nanorods.[8] Such quenching of PL also manifests itself in Au-PbS nanohybrid composites, compared to pure PbS quantum dots (QDs).[9] Despite the above-mentioned investigations, there are still several fundamental yet technologically important phenomena that required to be understood. One of the phenomena is energy transfer under intense excitation. Here, we report the observation of intensity-dependent enhancement of saturable absorption in a thin film of PbS-$Au_4$ nanohybrid composites. Such an intensity-dependent enhancement becomes significant at higher excitation irradiance due to resonant energy transfer through an irradiance-dependent channel: Auger recombination.

PbS QDs were synthesized by a previously reported solution method.[10,11] In brief, 0.28 g $PbCl_2$ and 5 ml oleylamine were introduced to a three-necked round-bottom flask at room temperature. The reaction system was sealed, vacuumed and then purged by $N_2$ for one hour to remove oxygen and moisture. The solution was then heated to 90°C to form the Pb-oleylamine complex. An hour later, 2.5 ml oleylamine solution of sulfur (21 mg) was quickly injected into the above solution. The resulting solution was heated to 210 °C and aged at that temperature for one hour. After the solution cooled down to room temperature, methanol was added to precipitate PbS QDs followed by centrifugation. The solid product so-obtained could disperse well in toluene. The preparation of PbS-Au$_4$ nanostructures was done by the following procedure. 0.1 ml toluene solution of 0.2 ml HAuCl$_4$ and 50 mM tetraoctylamine bromide was added to 2 ml of 50 mM dodecylamine. The resulting gold mixture was stirred at room temperature overnight. At the same time a toluene solution of PbS QDs was diluted until the absorbance of the solution at 500 nm was about 1.15. This was to ensure that the PbS concentration was kept the same in all experiments. 4 ml diluted solution of PbS QDs was then introduced to a three-necked round-bottom flask. The reaction system was sealed, vacuumed and purged by N$_2$ for one hour. The solution was then slowly heated to 40 ˚C. 30 min later, the stock solution of gold was injected into the PbS solution. The reaction was allowed to proceed at that temperature for one hour. The reaction was then stopped and the solution was centrifuged at 16000 rpm for 10 min. The thin films of PbS QDs and PbS-Au$_4$ nanohybrid composites were produced by spin coating at room temperature, and were utilized for the measurements presented below.

Figures 1(**a**) and 1(**b**) display the high-resolution transmission electron microscopic (HRTEM) images of PbS QDs and PbS-Au$_4$ nanohybrid composites, respectively. The HRTEM images clearly illustrate the monodispersity of PbS QDs which have a diameter of ~12 nm, while the size of Au particles in PbS-Au$_4$ is ~ 3 nm. Since the size of PbS QDs is smaller than the exciton Bohr radius of PbS (18 nm), the strong quantum confinement effect should be expected. Figure 1(**c**) shows the UV-Vis-NIR spectra of PbS QDs and PbS-Au$_4$ nanohybride composites. The absorption due to the surface plasmon of Au nanoparticles is not apparent in the spectrum of PbS-Au$_4$

nanohybrid composites, different from the results of Shi *et al.*[9] The difference could arise from the smaller size (~3 nm) for the Au nanoparticles, and the lower Au:PbS ratio in the nanohybrid composites, causing the Au surface plasmon absorption to be submerged in the more prominent PbS absorption. In comparison with the absorption spectrum of pure PbS QDs, however, there is pronounced difference, in particular, disappearance of a shoulder around 400 nm (or ~3 eV) which provides evidence for the influence of 3-nm-sized Au nanoparticles in the nanohybrid composites. It should be noticed that the absorption spectra in Fig. 1(**c**) were measured as PbS QDs and PbS-Au$_4$ nanohybrid composites were in form of thin films, while the spectra in Ref. 11 were obtained from their solution in toluene.

Figure 1(**d**) depicts both measured one-photon absorption spectrum and calculated transition bands of PbS QDs. Based on the theoretical model reported in Ref. 12, we find that the measured absorption spectrum can be fitted well by taking into consideration several excitonic transitions with Gaussian broadening which correspond to the size dispersion of QDs. The lowest excitonic transition ($E_{10}$) is peaked at 0.74 eV for PbS QDs of 12-nm diameter. A blueshift of 330 meV is found in comparison with the band-gap energy of bulk PbS crystal. Since our photon energy used here is ~1.59 eV, the mechanism primarily responsible for saturable absorption is band filling by one-photon excitation onto excitons of $E_{20}$ and $E_{21}$ centered at 1.71 eV and 2.44 eV, respectively.

The standard time-resolved transient absorption measurement (or degenerate pump-probe experiment) was carried out using 180-fs laser pulses at 1 kHz repetition rate with lower average power to minimize accumulative thermal effects. The laser pulses were generated by a Ti: Sapphire regenerative amplifier system (Quantronix, Titan seeded by Quantronix, IMRA). The pump and probe beams were focused onto the thin film with a minimum beam waist of ~ 30 μm. Previously, we reported [11] the observation of reverse saturable absorption (RSA) at 700 nm, which could be attributed to the dominance of free-carrier (or intraband) absorption[13,14] and nonlinear scattering mechanisms[14]. Saturable absorption, particularly due to the surface plasmon of Au nanoparticles, was also evident as the magnitude of RSA was reduced in PbS nanohybrid composites in comparison with pure PbS QDs[11]. To make saturable absorption be predominant, the excitation wavelength used here is altered to 780 nm whereby RSA becomes insignificant as discussed in the following. The origin of alteration from RSA to saturable absorption is unclear. We speculate that more hot excitons of 2P(h)-2P(e) (centered at $E_{21}$ = 2.44 eV) might be excited with excitation wavelength at 700 nm (or 1.77 eV) which is closer to the center of $E_{21}$. Due to the large broadening, as presented in Fig. 1(**d**), there should be highly probable for the hot carriers to further absorb photons, giving rise to free-carrier absorption (or intraband) absorption.

At 780 nm, Fig. 2 shows a positive change in the transmission which results from photobleaching or saturable absorption. At excitation irradiances of 27 GW/cm$^2$ or below, the PbS QDs and PbS-Au$_4$ nanohybrid composites exhibit the nearly identical responses, which imply that the Au nanoparticles make no contribution. As the laser irradiance is increased to 78 GW/cm$^2$, however, the PbS-Au$_4$ nanohybride composites demonstrate stronger photobleaching than the PbS QDs. To evaluate the magnitude of saturable absorption, we compare the nonlinear absorption coefficient ($\alpha_2 = -\alpha/I_{sat}$) of PbS-Au nanohybrid composites with that of PbS QDs by the following expression:

$$\alpha_2^s = \alpha_2^r \times \left(\frac{|\Delta T|_s}{|\Delta T|_r}\right) \times \left(\frac{1-R_r}{1-R_s}\right)^3 \times \left(\frac{L_r}{L_s}\right) \times \frac{\exp(\alpha_s L_s)}{\exp(\alpha_r L_r)}, \tag{1}$$

where subscripts $s$ and $r$ denote the PbS-Au$_4$ nanohybrid composites and the PbS QDs, respectively, $L$ is the interaction length of pump and probe beam over the sample, $\alpha$ is the small-signal absorption coefficient, $I_{sat}$ is the saturation intensity, and $R$ is the surface reflectance. From Fig. 2, an enhancement of nearly fourfold in the $\alpha_2$ value is inferred for the PbS-Au nanohybrid composites in comparison with the PbS QDs. The absolute $\alpha_2$ value of PbS QD-doped glasses was reported to be 0.75 cm/GW by 15-ps laser pulses.[15,16] With this reference value, one estimates the absolute $\alpha_2$ value of PbS-Au$_4$ nanohybrid composites to be as large as ~ 2.9 cm/GW.

This enhancement at higher excitation irradiance is attributed to resonant energy transfer from PbS QDs to Au nanoparticles by Auger recombination, as illustrated in Fig. 3. Upon on one-photon excitation, electrons are promoted to one of the states in the conduction band of PbS QDs. Due to higher excitation, there is more than one electron per QD in the excited states, triggering Auger recombination in which an electron recombines with a hole and the excess energy is used to promote another excited electron to a higher excited state. The Auger process has been demonstrated to be more efficient due to enhanced carrier-carrier interaction in the confined space of QDs in comparison with bulk materials.[17] The energy of the highly excited electron is then transferred to the Au nanoparticle since it is nearly resonant with the surface plasmon in the range of 2.2~3.5 eV. As a consequence, surface plasmonic effects in the Au nanoparticle occur, providing extra photobleaching to the transient transmission and leading to the intensity-dependent enhancement.

To gain more insights into this dynamic process, we fit the pump-probe data with the two-exponential component model, corresponding to the two relaxation processes: the fast (2 ~ 9 ps, depending on excitation irradiance) and the slow (~110 ps) decay components. The details of fitting parameters are tabulated in Table 1. The irradiance dependence of the fast component provides clear evidence for occurrence of Auger process while the slow decay is due to carrier recombination in PbS QDs.

In conclusion, we have presented the synthesis and time-resolved nonlinear transmission measurements of PbS-$Au_4$ nanohybrid composites. The nonlinear absorption coefficient $α_2$ of saturable absorption in PbS-$Au_4$ nanohybrid composites is enhanced by nearly fourfold in comparison with that of pure PbS QDs at excitation irradiances of 78 GW/cm$^2$, while no enhancement is observed at 27 GW/cm$^2$ or below. This irradiance-dependent enhancement is attributed to excitation of surface plasmon through resonant energy transfer from PbS QDs to Au nanoparticles by Auger recombination.

**Table 1.** Fitting parameters based on double-exponential decay: $\Delta T/T = A_1\exp(-t/\tau_1) + A_2\exp(-t/\tau_2)$.

| Sample | Irradiance (GW/cm$^2$) | $A_1$ | $\tau_1$ (ps) | $A_2$ | $\tau_2$ (ps) |
|---|---|---|---|---|---|
| PbS QD or PbS-Au$_4$ | 8.9 | $5.8 \times 10^{-6}$ | 9 | $5.9 \times 10^{-6}$ | 110 |
| PbS QD or PbS-Au$_4$ | 27 | $1.0 \times 10^{-5}$ | 6 | $8.3 \times 10^{-6}$ | 110 |
| PbS QD | 78 | $4.0 \times 10^{-5}$ | 2 | $1.0 \times 10^{-5}$ | 110 |
| PbS-Au$_4$ | 78 | $5.0 \times 10^{-5}$ | 2 | $2.0 \times 10^{-5}$ | 110 |

**Figure Captions:**

Figure 1 (Color online) HRTEM images of starting PbS QDs (**a**) and PbS-Au$_4$ nanohybrid composites (**b**), UV-VIS-NIR spectra of PbS QDs and PbS-Au$_4$ nanohybrid composites (**c**); calculated transition energies ($E_{nl}$) of PbS QDs and their Gaussian broadening fitting (**d**). $E_{nl}$ is defined as $E_{nl} = E_{nl}^{c} - E_{nl}^{h}$, where $E_{nl}^{c} = \hbar^2 \xi_{nl}^2 / 2R^2 m_e$, $E_{nl}^{h} = -E_h - \hbar^2 \xi_{nl}^2 / 2R^2 m_h$, R is the radius of QDs, $\xi_{nl}$ is the $n$th root of the spherical Bessel function of $l$th order, $J_l(\xi_{nl}) = 0$, and m$_{e,h}$ is the effective mass of electron (or hole).

Figure 2 (Color online) Time-resolved transient change in transmission ΔT/T measured in PbS QDs and PbS-Au$_4$ nanohybrid composites with 180-fs pulsed excitation at 780 nm.

Figure 3 (Color online) Illustration of resonant energy transfer by Auger process.

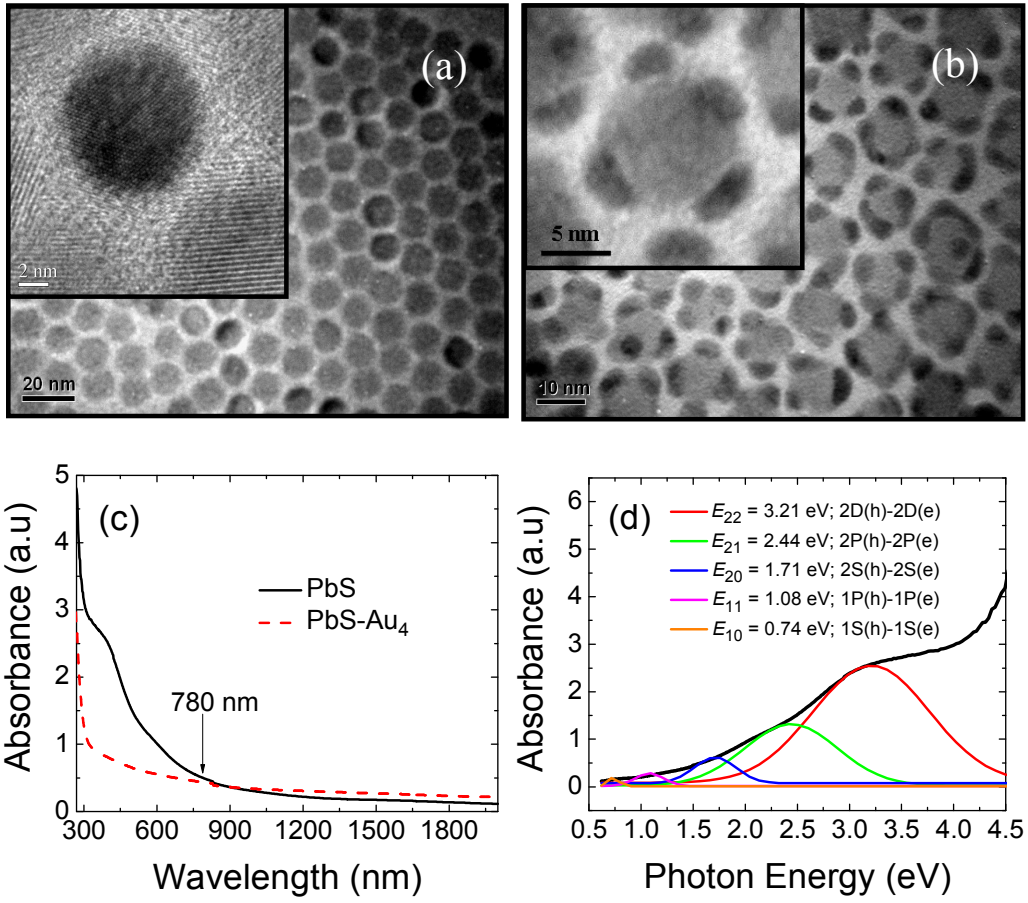

**Figure 1**.

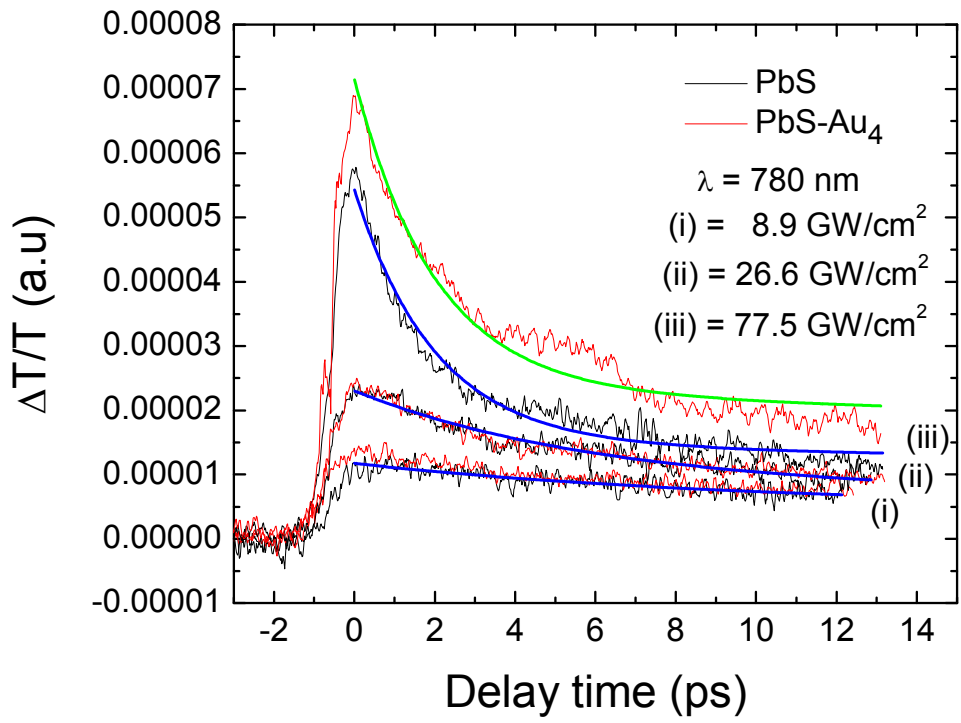

**Figure 2.**

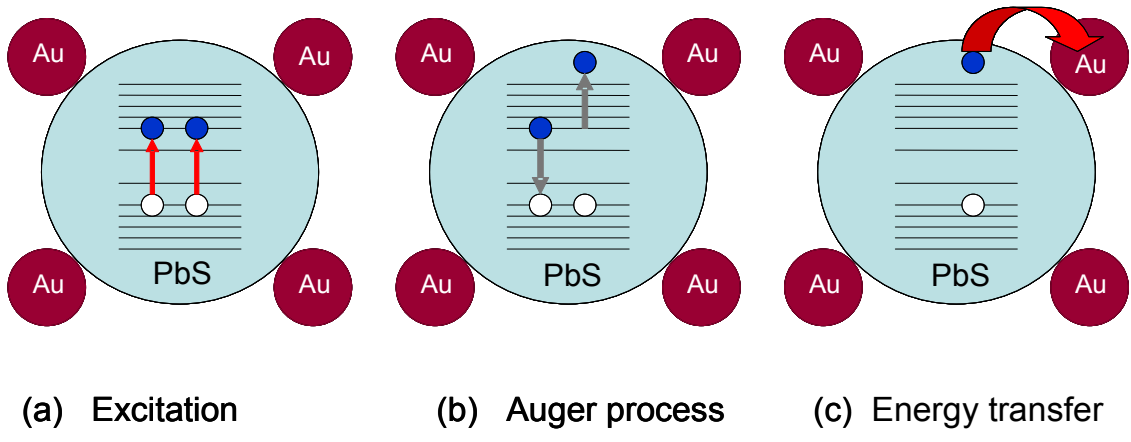

**Figure 3.**